\pdfoutput=1
\documentclass{article}
\usepackage{lscape}
\usepackage{longtable}
\usepackage[normalem]{ulem}
\usepackage{multirow}
\usepackage{indentfirst}
\usepackage{slashbox}
\usepackage{pict2e}
\usepackage[algoruled,linesnumbered]{algorithm2e}
\usepackage{graphicx}

\begin{document}

\title{A Distributed Transportation Simplex Applied to a Content Distribution Network Problem}

\author{Rafaelli de C. Coutinho \and L\'ucia M. A. Drummond \and Yuri Frota \thanks{\{rcoutinho,lucia,yuri\}@ic.uff.br} \\
Institute of Computing -- Fluminense Federal University, RJ} 




\maketitle

\begin{abstract}
A Content Distribution Network (CDN) can be defined as an overlay system that replicates copies of contents at multiple points of a network, close to the final users, with the objective of improving data access. 
CDN technology is widely used for the distribution of large-sized contents, like in video streaming.
In this paper we address the problem of finding the best server for each customer request in CDNs, in order to minimize the overall cost. We consider the problem as a transportation problem and a distributed algorithm is proposed to solve it. The algorithm is composed of two independent phases: a distributed heuristic finds an initial solution that may be later improved by a distributed transportation simplex algorithm.
It is compared with the sequential version  of the transportation simplex and with an auction-based distributed algorithm.  Computational experiments  carried out on a set of instances adapted from the literature revealed that our distributed approach  has a performance similar to its sequential counterpart, in spite of not requiring  global information about the contents requests. Moreover, the results also showed that the new method outperforms the based-auction distributed algorithm.
\end{abstract}

\section{Introduction}\label{sec:introduction}

A Content Distribution Network (CDN) can be defined as an overlay system that replicates copies of contents at multiple points of a network, close to the final users, with the objective of improving data access. Some problems need to be considered in order to implement and operate a CDN. 
At first, the servers need to be strategically located in the network, 
aiming to cover a wide range of potential clients. This problem is called the Server Placement Problem 
(SPP)\cite{ref2}. Once servers are placed, and given a short-term forecast (say, for the next hours)
of the clients requests for contents, it is necessary to decide how contents should be replicated among the servers in order to minimize the expected
access time. This second problem is called the Replica Placement Problem (RPP) \cite{ref10}. 
The last one, called the Request Routing System Problem (RRSP) consists in, given the client requests and the configuration of the CDN (server/link capacities, latency, costs, etc), finding the best assignment of each request to a server that holds a copy of the required content. Remark that, while it is usual that most requests will be really assigned to
a single server (the strict meaning of the word  ``assignment'' in the optimization community), the problem definition allows splitting the demand of a request on different servers. This work proposes viewing and solving the RRSP as a Transportation Problem (TP). The Transportation Problem (TP) is a classical optimization problem \cite{networkflows}
that deals with sources where a supply of some items are 
available and destinations where the items are demanded. The objective  
is finding a minimum cost way of transporting the items. 
Given its vast range of applications, the TP has obviously been extensively
studied. 
The traditional sequential approach can only tackle a distributed problem after all data is collected
in a central node, then the solution is broadcast over the network.
However, it becomes increasingly harder to use this sequential approach, as 
the volume of data to be collected over a large wide area network increases. The difficulty lies
in maintaining the data updated in a highly dynamic environment. In fact,
the exponential growth of the Internet over the last three decades could only be
sustained because some of its key protocols were designed as distributed algorithms.

In  this work,  we propose a distributed version of the Transportation Simplex, an algorithm that was devised by Dantzig as a specialization of the Simplex algorithm for general linear programming problems \cite{Dantzig}. Although the Transportation Simplex is still the most usual algorithm for the sequential TP, to the best of our knowledge, no effort was made to obtain
 solution for its distributed version, where the sources and destinations 
are nodes of an actual network and the problem data is distributed among them.

In this paper we also compare the proposed algorithm with an auction-based algorithm. The auction algorithm is a parallel relaxation method for solving the classical assignment problem and  has been generalized to solve linear transportation problems in \cite{bertsekas}. The algorithm operates like an auction, there is a price for each object, and at each iteration, unassigned persons bid simultaneously for objects thereby raising their prices. Objects are then awarded to the highest bidder.  Although it has been proposed for shared memory multiprocessors, we adjusted it so that it also executed in distributed environments.  


The remaining of the paper is organized as follows. Section \ref{rel_work} provides an overview of related works that solve the transportation problem in parallel. In Section \ref{sec:formulation} we show the modelling of the RSSP as a transportation problem. 
In Section \ref{sec:auction} the developed distributed auction algorithm is shown.
The proposed distributed algorithms, both the one used to obtain an initial solution as the distributed  transportation simplex,  are presented in Section \ref{sec:distributed}.  
Section \ref{cap:resultado} discusses the implementations  and report the results of computational tests. Finally, Section \ref{conclusao} concludes the paper.


\section{Related Work}
\label{rel_work}
In the related literature,  two well known approaches are presented to solve  the transportation problem in parallel. The first one consists of parallel versions of the network simplex algorithm. Those algorithm solves
the minimum cost network flow problem, that includes the transportation problem as a particular case. The second approach considers auction algorithms to solve linear transportation problems.
  
In \cite{ref8}, a  primal network simplex algorithm, that applies decomposition of the original problem into subproblems, is proposed. Three variants of this parallel algorithm are presented. In the first one, each  process executes the same algorithm over a  different subproblem. Processes exchange parts of subproblems, periodically,   in order to eliminate arcs that connect subproblems of different processes, called {\em across arcs}.  In the second strategy, additionally to the procedure proposed previously, some processes are dedicated  uniquely to perform pricing operations. Finally, in the third method, a process can interrupt another one,  to transfer  a subproblem when  an across arc appears in the solution.   
In all cases, the global  optimal solution is found when all local problems are solved and there are no candidate across arc to enter  the solution. Tests were performed in a shared memory multiprocessor. Almost linear speedups were obtained on
instances corresponding to multi-period network problems.

A parallel dual simplex algorithm for shared memory parallel computers is introduced in \cite{ref9}.  The traditional simplex (primal or dual) does not offer many possibilities for parallelism, because it moves from a feasible basic solution to another by performing one pivoting operation at a time.  Then, Thulasiraman {\em et al.} proposed a dual simplex method  that performs concurrent pivoting operations, by executing each of them over small subgraphs. This operation consists of traversing the spanning tree of the graph, corresponding to a basis, from leaves to root, identifying  clusters that can perform  concurrent pivoting.
Each process consists of a graph node and the communication between processes uses shared memory. Tests were run in the  multiprocessor BBN Butterfly. Good speedups were observed only for a small number of processors, adding more processors did not decrease computational times. This indicates that the test instances only allowed a limited number of concurrent pivots.

In \cite{miller2} the authors presented a similar approach by introducing a synchronous parallel primal simplex algorithm for transportation problems on shared memory parallel computers. The method also performs concurrent pivoting operations by distributing the cost matrix across the machine with each processor storing a contiguous block of rows. 
Comparisons were made with parallel assignment problem algorithms but there was no clear winner. 

Later, a parallel primal simplex method for minimum cost network flow  problem is also proposed in \cite{ref3}. A greater degree of parallelism is achieved by also breaking down the pricing (not only the pivots) into different processors. A monitor is used to synchronize parallel processes and  to schedule tasks to processors. Those tasks  can  (i) select an arc and make the pivoting, (ii) update the dual variables, or (iii) perform the pricing operation. Tests were  executed on a shared memory multiprocessor, Sequent Symmetry S81 with 20 processors and 32 Mbytes of shared memory. For the instances tested, an average speedup of $8.26$  was reached.

A recent review of parallel simplex methods is presented in \cite{ref7}. According to it, there is no parallelization of the simplex method that offers significant improvement in performance for  large and sparse LP problems. However, some implementations of parallel solvers applied to dense or specific classes of linear programs were successful.

Concerning  auction-based parallel algorithms, Bertsekas and Casta\~non \cite{bertsekas}  proposed the converting of the transportation problem into an assignment one, intending to use a previously introduced auction algorithm for the assignment problem with small modifications to exploit the special structure of the network. Later, they discussed  about  parallel implementations of the auction algorithm in shared memory multiprocessors   with several degrees of synchronization among the processes for the assignment problem \cite{bertsekas2}. More recently, a particular case of the assignment problem, that deals with the lack of global information and limited communication capability, is solved by a distributed auction algorithm  that maximizes the total assignment within a linear approximation of the optimal solution \cite{zavlanos}. In \cite{park}, a novel variation of the assignment problem   is tackled by an  auction algorithm  that considers a  hierarchy of decision makers.

\section{Modeling the RSSP as a Transportation Problem}
\label{sec:formulation}

Let $G = (V, A)$ be a complete bipartite graph, where the vertex set $V$ is 
divided into subsets $V_1$ with $n$ sources and $V_2$ with $m$ destinations. The arc $(i,j)$ from source $i$ to
destination $j$ has a cost $c_{ij}$. Each source $i \in V_1$ is associated with a capacity $b_i$, while each
destination $j \in V_2$ is associated with a demand $d_{j}$. It is assumed that capacities and demands are balanced,
i.e., $\sum_{i \in V_1} b_i = \sum_{j \in V_2} d_j$; if this is not the case, an artificial vertex can be introduced to
absorb the excess of capacity or demand. The Transportation Problem (TP) consists in solving the following linear program: 
\begin{eqnarray}
\min & \displaystyle \sum_{i\in V_1} \sum_{j\in V_2} c_{ij} x_{ij} \hspace{.5cm}
\label{obj2} \\
s.t. &  \displaystyle \sum_{i\in V_1} x_{ij} = d_j, & \forall j \in V_2, \label{rest11} \\
& \displaystyle \sum_{j \in V_2} x_{ij} = b_{i}, & \forall i \in V_1, \label{rest21} \\
& x_{ij}  \geq 0,& \forall i \in V_1,\ j \in V_2,\label{rest31}
\end{eqnarray}
where variables $x_{ij}$ represent the flow from $i$ to $j$.
Remark that any of the equalities in (\ref{rest11}-\ref{rest21}) is implied by the remaining
$m+n-1$ equalities and can be removed. 

To cast the RRSP as a TP, some considerations are made about typical CDNs:
\begin{itemize}
	\item A CDN is composed by a set $I$ of servers, spread around a large geographical area, that can communicate using the underlying Internet structure.
	\item A CDN hosts a set $C$ of distinct contents. At a given moment, a server $i \in I$ has a subset $C_i$
	of contents, determined by the last solution of the Replica Placement Problem. Alternatively, it can be said that
	for each $c \in C$, $I_c \subseteq I$ is the subset of servers that have $c$.
	It is usual that only a single central server (which is actually a datacenter) contains a copy of all contents.
	\item The servers have a limited amount of outgoing bandwidth available. The bandwidth of server $i \in I$ is given by $b_i$.

	\item At a given moment, there is a set $J$ of client requests. A request $j \in J$ is associated with a content $c(j) \in C$ and with a required bandwidth of $d_j$. The value of $d_j$ is related to Quality of Service issues. Sometimes $d_j$ is defined
	as a function of the content $c(j)$. $J(C_i) \subseteq J$ is the subset of requests for a content in $C_i$.
	
	\item There is a subset $K \subseteq I$ of the servers (called {\em client servers}) that not only hold contents, they are
	also associated to the client requests. In other words, each request $j \in J$ is associated with a server $k(j) \in K$. When content $c(j)$ belongs to $C_{k(j)}$, the request can possibly be attended by $k(j)$ itself. Otherwise, it must be redirected to servers that contain $c(j)$. In any case, the total consumption of bandwidth of the servers attending $j$ is $d_j$.

	\item It is assumed w.l.o.g. that there exists at most one request in server $i$ for a content $c$, so a request $j \in J$ can
	be alternatively identified by the pair $(k(j),c(j))$.

  \item The CDN administrator can define the communication cost $c_{i_1i_2}$ between two servers
  $i_1$ and $i_2$ in $I$, per unit of traffic, based on parameters like distance or latency. If $i_1=i_2$, the cost is zero. This means that fully attending a demand $j$ by $k(j)$, if it is possible, costs zero. But this would still consume the bandwidth of $k(j)$ by $d_j$ units.
\end{itemize}

Given the above considerations, the RRSP can be modeled as the following linear program:
\begin{eqnarray}
\min & \displaystyle \sum_{j\in J} \sum_{i\in I_{c(j)}} c_{ik(j)} x_{ij} \hspace{.5cm}
\label{obj} \\
s.t.&  \displaystyle \sum_{i\in I_{c(j)}} x_{ij} = d_j, & \forall j \in J, \label{rest1} \\
& \displaystyle \sum_{j \in J(C_i)}  x_{ij} \leq b_{i}, & \forall i \in I, \label{rest2} \\
& x_{ij}  \geq 0,& \forall j \in J,\ i \in I_{c(j)},\label{rest3}
\end{eqnarray}
where variables $x_{ij}$ are the bandwidth that server $i$ uses to attend
demand $j$.

By associating $V_1$ to $I$ and $V_2$ to $J$, the relation between (\ref{obj2}-\ref{rest31}) and (\ref{obj}-\ref{rest3})
becomes evident. However, in order to obtain a perfect reduction of the RRSP to a TP, some details must be fixed:
\begin{itemize}
	\item The bipartite graph induced by (\ref{obj}-\ref{rest3}) is not complete. The solution is adding the missing arcs, between requests and servers that do not contain the required content, with suitable large costs. We remark that even when many artificial arcs are introduced, this has a small impact on the performance of the distributed
algorithms that will be proposed. This happens because most of those arcs can be handled implicitly and they are completely ignored as soon as a first truly feasible solution is found.
  \item Converting (\ref{rest2}) to equality can be easily done by adding an artificial demand of $\sum_{i \in I} b_i - \sum_{j \in J} d_j$ units. If this quantity is negative, then the problem is infeasible.
Anyway, while adding an artificial demand is trivial for a sequential TP algorithm, it is not so obvious when designing a distributed algorithm, since it requires a global data that is not readily available in the beginning of the algorithm.
\end{itemize}


\section{The auction algorithm for the TP}\label{sec:auction}

The auction algorithm is a parallel relaxation method for solving the classical assignment problem.
The algorithm operates like an auction whereby  unassigned persons bid simultaneously for objects 
by raising their prices. Each object $j$ has a price $price_j$, with the initial prices being zero.
 Prices of the objects  are adjusted upwards as persons bid for the best objects, that is the object for which the corresponding benefit minus the price is maximal. Only persons without an object submit a bid, and objects are awarded to their highest bidder. When the prices of some of the assigned objects become  high enough, unassigned objects become attractive and  receive new bids. Thus, the algorithm executes several bidding and assignment steps towards  a full assignment of persons to objects \cite{bertsekas}.

The transportation problem is converted into an assignment problem by creating multiple copies of similar persons or objects for each source or destination, respectively.
Two objects are called similar if every person to whom they can be assigned considers them as equally valuable, while two  persons  are named similar if  they assign the same value to every object. 
The  objective of the problem is to maximize the costs of the flows between sources and destinations. 
The  bidding  and  the  assignment  phases  of the  auction  algorithm  are  highly parallelizable with shared memory. The bidding  can be accomplished simultaneously for  all persons. Similarly, the assignment  can be accomplished simultaneously for all objects.

Considering that  the servers of a CDN  communicate through a network and do not have a shared memory,  the algorithm by Bertsekas and Casta\~{n}on had to be  adapted to execute in a distributed way.   Here the sharing of  prices and bids  is  accomplished through message exchanges among sources and destinations. The Algorithm \ref{atp}, named {\sl AuctionTP},  presents our  proposal for a  based-auction distributed algorithm for the TP. 
Each source accomplishes the bidding phase, by  sending flows and bids to each destination not yet  completed served by it, and by  calculating  new flows and bids according with the answers received from these destinations.  Each destination receives all the  sent bids, updates the flows and prices of its requests and sends an acknowledge message to all servers containing the new prices, flows  and the corresponding sources responsible for serving it.
The procedure of updating bids (\ref{a}), offering flows (\ref{b}), prices (\ref{d}),  accepted flows (\ref{d}) and  $\varepsilon$ (\ref{f} and \ref{e})  are accomplished according with the rules proposed by  \cite{bertsekas}.
 The algorithm  terminates with an optimal solution provided that the transportation problem is feasible and $\varepsilon < 1/min\{m, n\}$, where $n$ and $m$ are the number of sources and  destinations, respectively. 
Remark that as  RRSP consists of a minimization problem we  multiplied  each cost $c_{ij}$ by $(-1)$ and added  the highest cost among  all of them to each one.

\begin{algorithm}[htb]
\SetAlgoNoLine

\textbf{Initialization}

\textbf{Initial Assignment:}  $flow_{aj} := d_j$, $\forall j \in J$, where $a$ is an artificial source;\\
\For{every source $i$ that offers a content required by destination $j$}{ 
     $bid_{ij} :=0$;\\
     $offer\_flow_{ij} :=0;$\\
     $price_{ij} := 0 $;\\
     $flow_{ij} := 0 $;\\
}
$\varepsilon := (max_{i \in I, j\in J}\{c_{ij}\} \times min\{n,m\})/2$; \label{f}\\

\vspace{0.5cm}
\textbf{Bidding Step executed by source $i$} 

\For{each destination $j$ not totally attended by source $i$ that requires a content in $C_i$}{
   \textbf{update} $bid_{ij}$ ($flow_{ij}$, $price_{ij}$, $c_{ij}$, $\varepsilon$);\label{a}\\
   \textbf{update} $offer\_flow_{ij}$ ($flow_{ij}$, $price_{ij}$, $c_{ij}$);\label{b} \\
   \textbf{send} message($bid_{ij}$, $offer\_flow_{ij}$, $i$) to destination $j$; \label{c}
   } 
\textbf{Upon receiving} message ($ACK$, $local\_price$, $local\_flow$, $local\_server$) from destination $j$:\\

\hspace{0.5cm}   \textbf{update} $price_{kj}$ and  $flow_{kj}$, $\forall k \in local\_server$;  \label{d}\\
\hspace{0.5cm}   \textbf{update}  $\varepsilon$;\label{e}\\

\vspace{0.5cm}
\textbf{Assignment Step executed by destination $j$}

\textbf{Upon receiving} message ($bid_{ij}$, $offer\_flow_{ij}$, $i$) from every source $i$ that keeps a  required content:\\ 
\hspace{0.5cm} $list$ :=  received messages  sorted by  decreasing order of bid;\\
\hspace{0.5cm} $local\_price : =  \emptyset$;\\
\hspace{0.5cm} $local\_flow := \emptyset $;\\
\hspace{0.5cm} $local\_server := \emptyset $; \\
\While  {(\textbf{not} full attended) or ($list \neq \emptyset $)}{
      $price\_flow\_received_i$ := first element removed from $list$;\\
      $local\_price := local\_price \cup bid_{ij}$ from $price\_flow\_received_i$;\\
      $local\_flow := local\_flow \ \cup $ updated $offer\_flow_{ij}$ from $price\_flow\_received_i$;\\ 
      $local\_server := local\_server \cup i$ from $price\_flow\_received_i$;\\
      }
\textbf{send} ($ACK$, $local\_price$, $local\_flow$, $local\_server$) to every source $k \in I$;

\caption{Auction Algorithm for the TP}
\label{atp} 
\end{algorithm} 

\section{Distributed Algorithm}\label{sec:distributed}


\subsection{Sequential Transportation Simplex} \label{subsec:sequential}

The Transportation Simplex algorithm needs an initial basic solution to start with.
The classical Northwest Corner method, described as Algorithm \ref{ncm}, is a constructive heuristic to obtain it. 
The method was initially devised as an easy to execute pencil-and-paper algorithm.
However, its solutions can be crude as the method completely disregards the costs, the next variable $x_{ij}$ that will receive
positive flow is arbitrarily chosen.
The related Minimum Cost method usually obtains a better initial solution by choosing a variable $x_{ij}$ that still can be increased with minimum $c_{ij}$. Anyway, the $n+m-1$ basic variables found by those methods define a tree in the bipartite graph $G$.

\begin{algorithm}[htb]
\SetAlgoLined
\KwIn{Integers $n$ and $m$, capacity and demand vectors $b$ and $d$}
\KwOut{Flow matrix $x$, boolean matrix $B$ identifying the basic variables}
$x=0$, $B=$ false, $s' = s$, $d' =d$\;
$i=1$, $j=1$\;
\Repeat{$d' = 0$} {
$\Delta = \min\{b'_i,d'_j\}$\;
$x_{ij}=\Delta$, $B_{ij}=$ true, $b'_i = b'_i - \Delta$, $d'_j = d'_j - \Delta$\;
\If{$d' \neq 0$ and $b'_i = 0$ and $d'_j = 0$} {
	$B_{i(j+1)}=$ true\;
}
\If{$b'_i = 0$} {
	$i = i+1$\;
}
\If{$d'_j = 0$} {
	$j = j+1$\;
}
}

\caption{Northwest Corner method}\label{ncm}
\end{algorithm}

The Transportation Simplex itself \cite{Dantzig} performs a sequence of iterations, where basic solutions (always defined by
trees in $G$) can be improved until they are proved to be optimal. Each source is associated with a dual variable $u_i$ and
each destination with a dual variable $v_j$. The construction of a dual solution 
uses the constraints $u_i +  v_j = c_ {ij}$, for each basic variable $x_{ij}$. One of those variables, say $u_1$, can be fixed to zero,
defining the root of the tree. Starting from that root, the remaining dual variables are determined in a unique way by propagating
the values over the tree. After that, the reduced costs of the non-basic variables can also be calculated as $\bar{c}_{ij} = c_{ij}- u_i - v_r$. If all reduced costs are non-negative, the solution is optimal. Otherwise, a variable with negative reduced cost is selected to enter the basis. This variable creates a cycle in the tree. Increasing the value of the entering variable by $\theta$ changes the values of the other variables along the cycle, by $-\theta$ and $\theta$, alternating. The entering variable is increased by the maximum possible value, which causes some variables to drop to zero. One of those variables is selected to leave the basis, which corresponds to another tree. A complete description of all those sequential algorithms can be found in
\cite{Bazaraa1990}.





\subsection{Distributed Approach}\label{subsec:distributed}

In the  distributed algorithms described next, each server in $I$ is associated to a process, that executes the same local algorithm, which consists of sending messages, waiting for incoming messages and processing. Messages can be transmitted  independently and arrive  after an unpredictable but finite delay, without error and in sequence.  Processes have distinct identities that can be totally ordered. Initially, a server $i$ only knows the requests $(i,c)$, $c \in C_i$. On the other
hand, it is assumed that a server knows which contents are present in each other server (possibly using a service similar to DNS) and the corresponding communication cost. The assumption is reasonable since content position change much less dynamically
than request data.

\subsubsection{Obtaining an Initial Solution}\label{subsubsec:initialdist}

\begin{figure}[h]
\centering
\begin{center}
\begin{tabular}{c}
\includegraphics[scale=0.9]{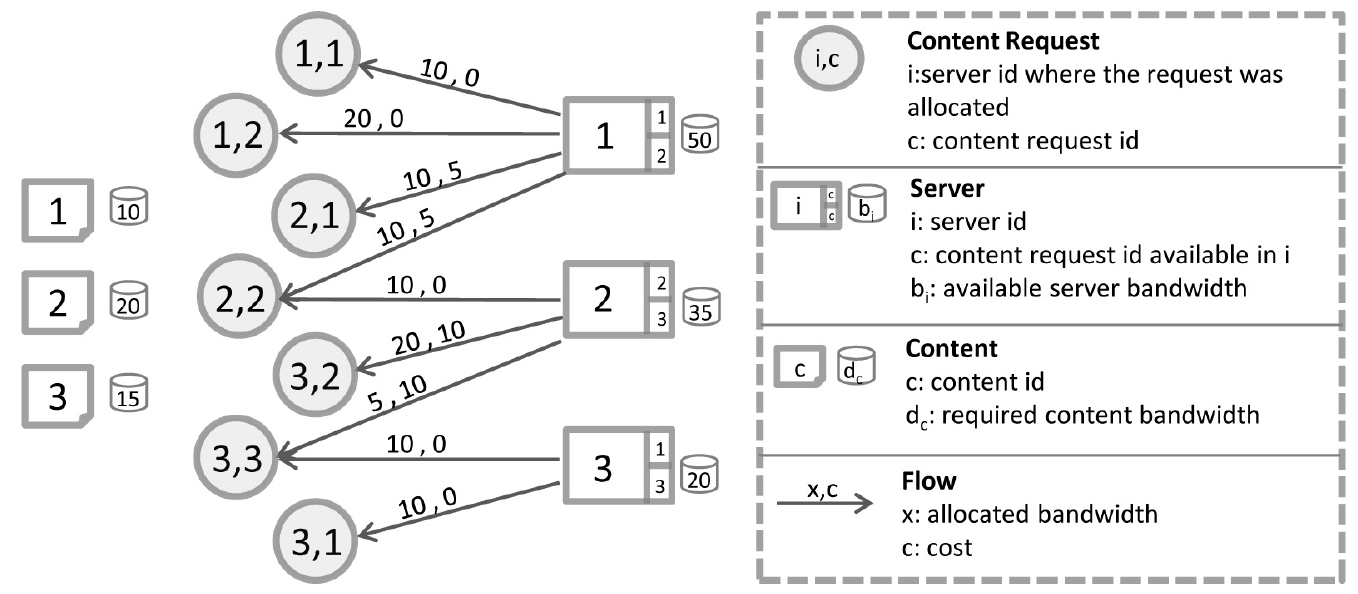} \\
\end{tabular}
\end{center}
\caption{Initial Solution}
\label{fig:1}
\end{figure}

{\sl DistInit} is a distributed constructive heuristic, inspired by the Minimum Cost method, but not necessarily equivalent to it.
In {\sl DistInit}, at first, each server tries to attend its local requests. If it is not capable of serving a local request, a message {\it Serve} is sent to the closest server that stores that content. Upon receiving this message, a server answers with an {\it ACK} or a {\it NACK} message depending on its bandwidth availability. When a server receives  a {\it NACK} or an {\it ACK} with partial attending, it sends another {\it Serve} message to the next closest server with the remaining required bandwidth.
When all requests are attended, the algorithm finishes.  See Figure \ref{fig:1} for  an example of an initial solution. Squares and circles represent servers and requests, respectively. The number inside the square is the server identification $i$, while the two numbers inside the circle of a request $j$ represent $(k(j),c(j))$. The edges indicate which servers attend each request.


\subsubsection{Distributed Transportation Simplex}\label{subsubsec:simplexdist}

The distributed transportation simplex ({\sl DistTS}) initiates in  the  server that has the smallest identification. It assigns  zero to its  dual variable $u_1$, calculates the dual variables $v$ corresponding to the requests supplied by it, and sends these values to the servers that hold those requests, through {\it Varv} messages. Upon receiving this message, each process calculates the dual variables of the other servers that have also contributed to attend a common request and sends these results through {\it Varu} messages.  This procedure is repeated, alternating the sending of {\it Varv} and {\it Varu} messages until all dual variables are calculated, and consequently the reduced costs.
Now, each server selects its most negative reduced cost edge and sends a message to the corresponding server, to introduce that non-basic variable in the solution.
Then,  messages are sent along the candidate cycles to select a basic variable to leave the solution. In the best case, this distributed algorithm can process all those cycles (i.e. make pivot steps) in parallel. However,  those candidate cycles may have intersections, which would cause inconsistencies if they are pivoted at the same time. So, when a conflict is detected, the cycle associated with the smallest reduced cost is chosen to continue, while the other is canceled.
For every pivot that is performed, it is
also necessary to re-calculate the dual variables. However, instead of recalculating all of them, only the ones that belong to the new subtree are updated. The new dual variables are propagated  through {\it Varu} and {\it Varv} messages and all procedures described above are repeated until it is detected that there is no edge with negative reduced cost. 

The steps of the Distributed Transportation Simplex can be summarized as follows. Those steps are illustrated by Figures 2 to 5, where an example of {\sl DistTS} execution is presented.

\begin{figure}[h]
\centering

\begin{tabular}{cc }
\includegraphics[scale=0.8]{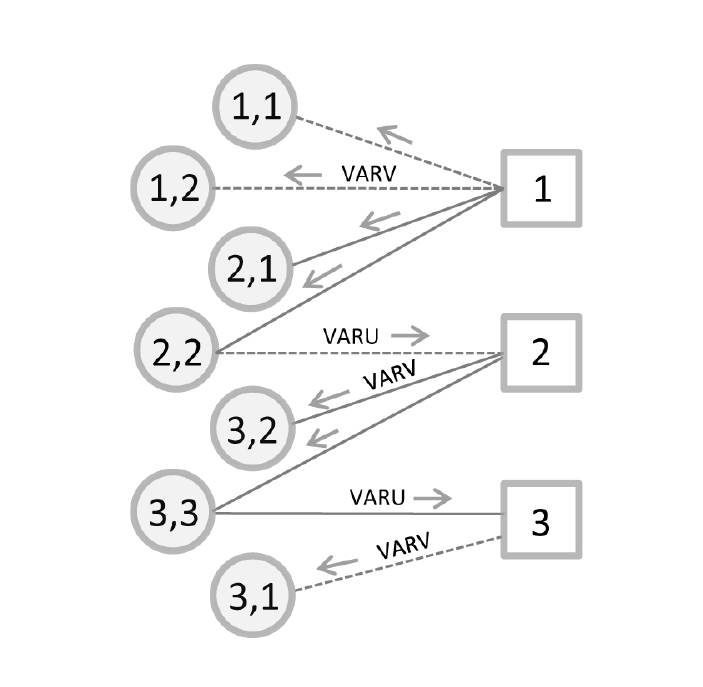}&  
\includegraphics[scale=0.7]{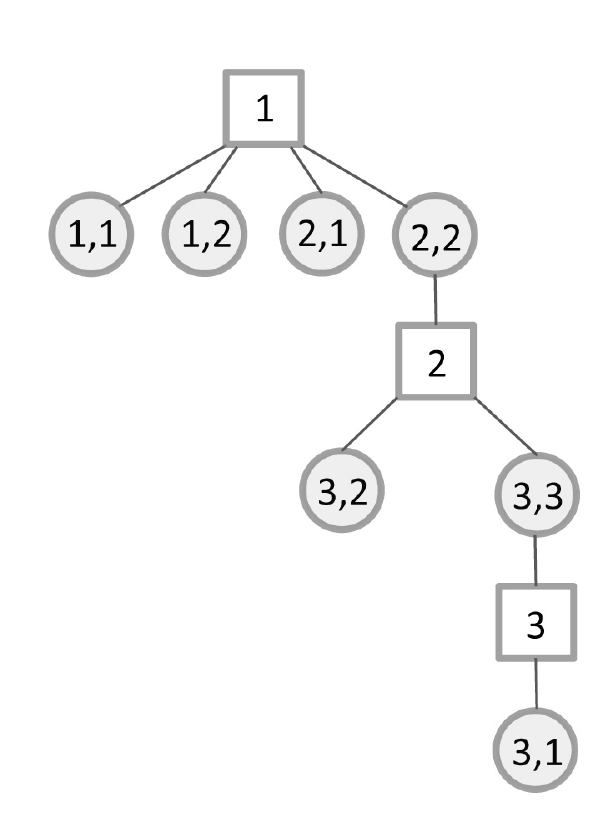} \\
{\bf {\footnotesize(a)}} &
{\bf {\footnotesize(b)}} \\
\includegraphics[scale=0.7]{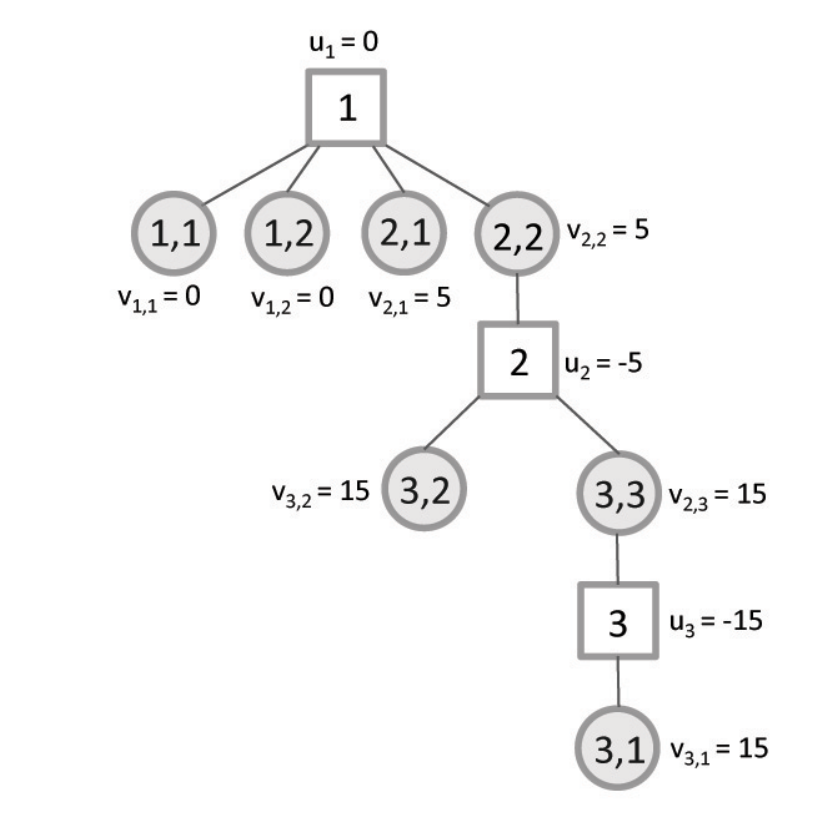}&  
\includegraphics[scale=0.7]{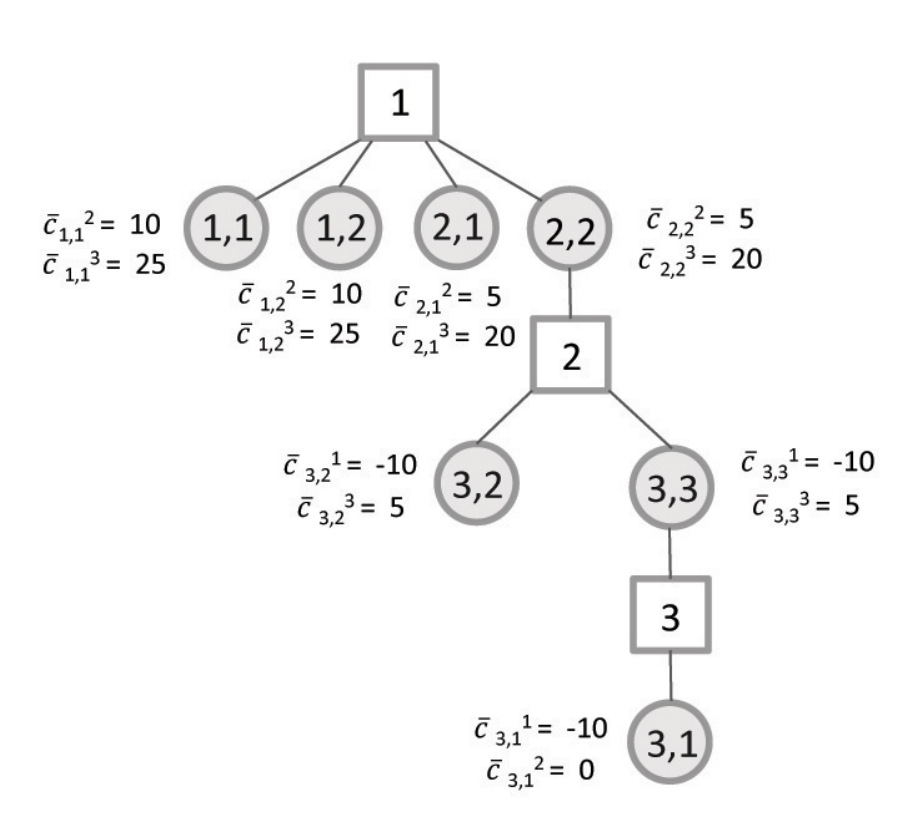} \\
{\bf {\footnotesize(c)}} &
{\bf {\footnotesize(d)}} \\
\end{tabular}

\caption{(a) {\it Varv} and {\it Varu} Messages; (b) Dual Solution Tree;  (c) Tree with Dual Variables; (d) Tree with Reduced Costs}
\label{fig:2}
\end{figure}

\begin{figure}[h]
\centering

\begin{tabular}{cc }
\includegraphics[scale=0.7]{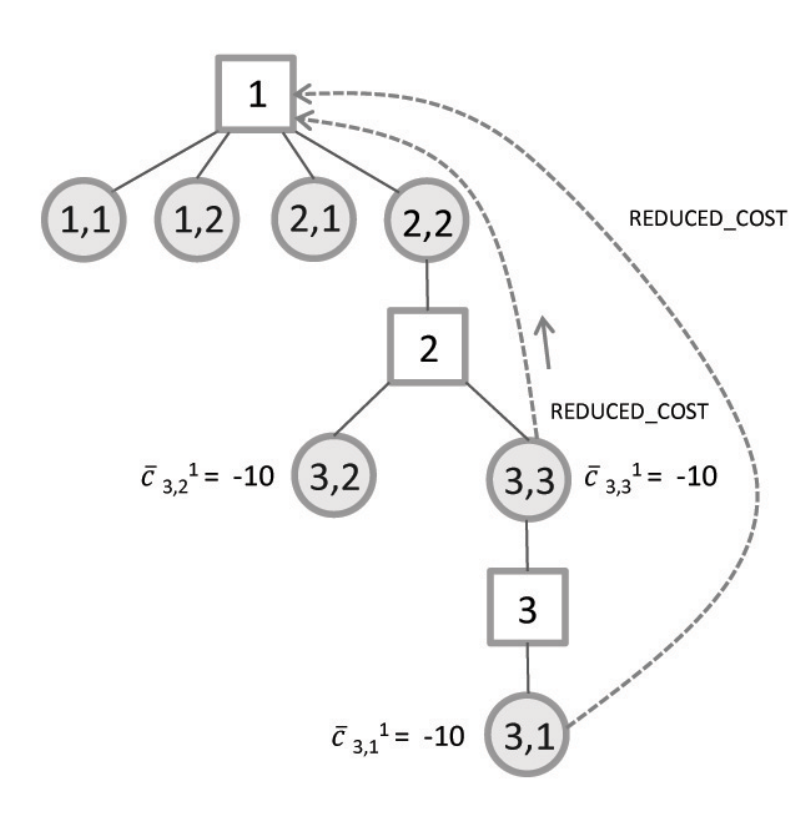}&  
\includegraphics[scale=0.7]{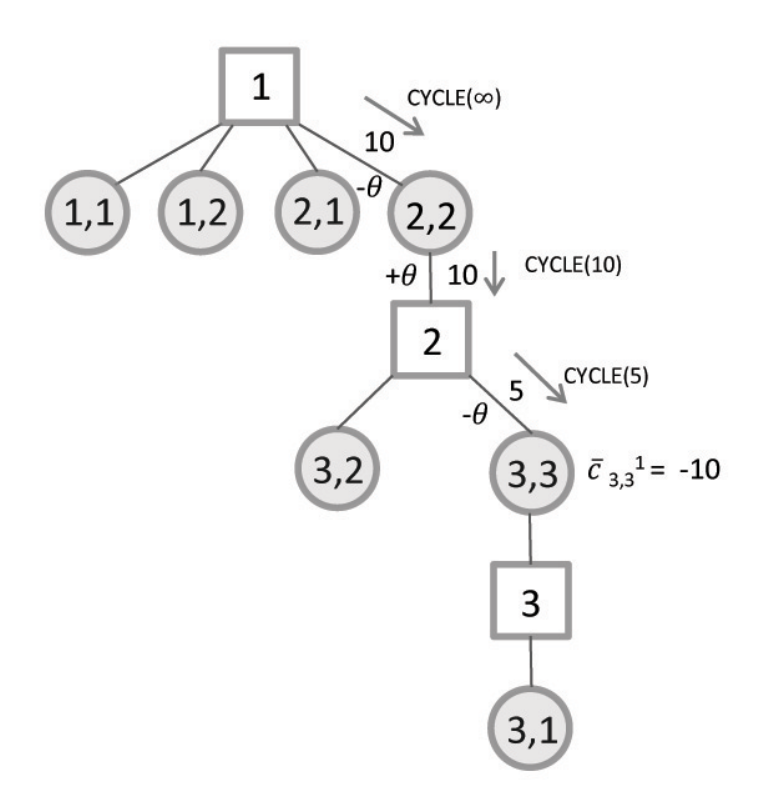} \\
{\bf {\footnotesize(a)}} &
{\bf {\footnotesize(b)}} \\
\includegraphics[scale=0.7]{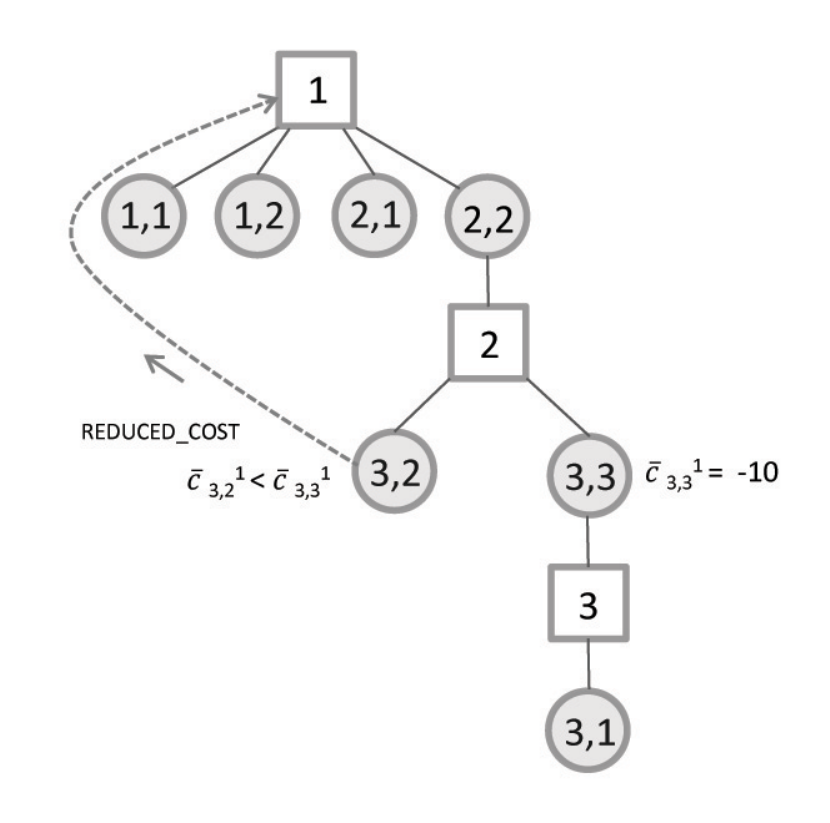}&  
\includegraphics[scale=0.7]{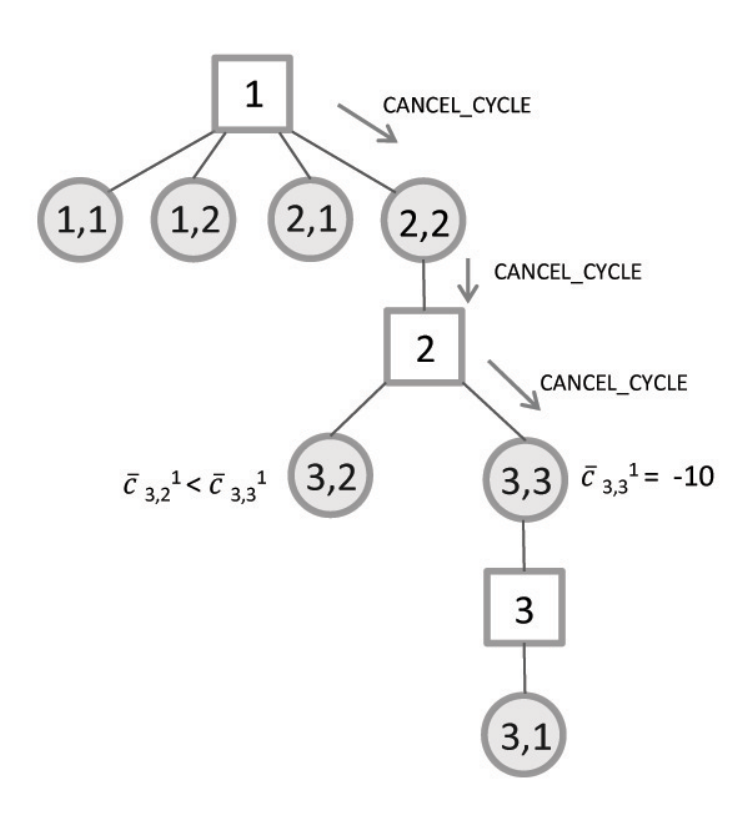} \\
{\bf {\footnotesize(c)}} &
{\bf {\footnotesize(d)}} \\
\end{tabular}

\caption{(a) Cycles corresponding to negative reduced costs; (b) Basic variable selection to leave the solution;  (c) Conflict detection; (d) Cycle canceling}
\label{fig:3}
\end{figure}

\begin{figure}[h]
\centering

\begin{tabular}{cc }
\includegraphics[scale=0.7]{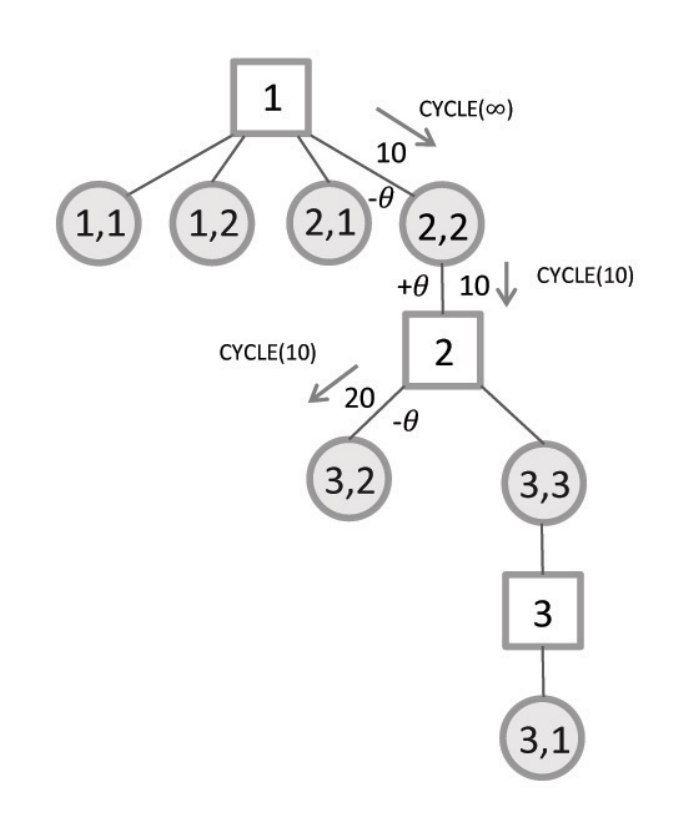}&  
\includegraphics[scale=0.7]{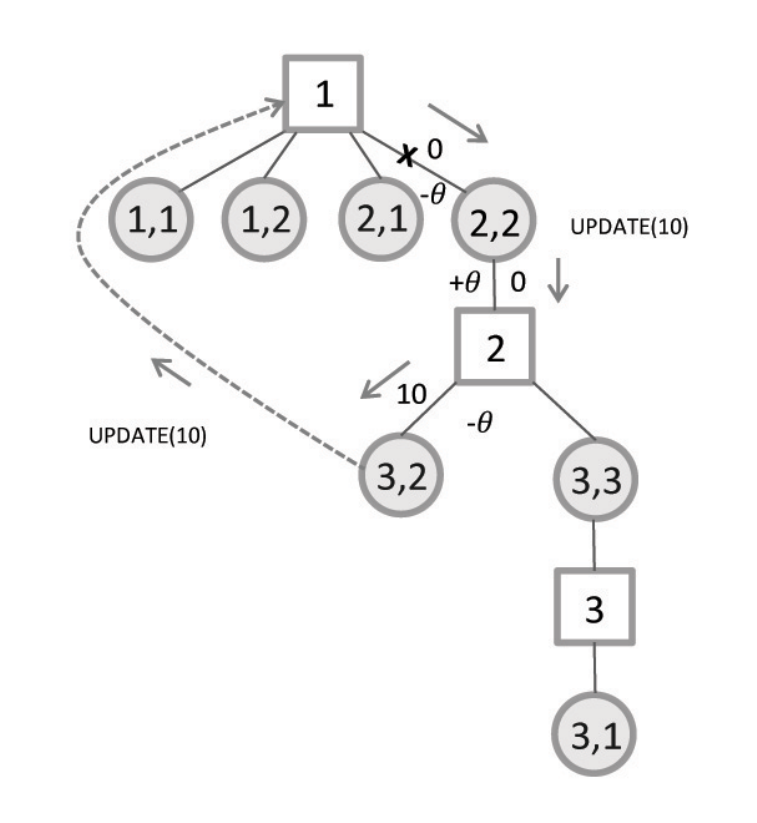} \\
{\bf {\footnotesize(a)}} &
{\bf {\footnotesize(b)}} \\
\end{tabular}
\caption{(a) Select other cycle;  (b) Updating the flows}
\label{fig:4}
\end{figure}

\begin{figure}[h]
\centering

\begin{tabular}{cc }
\includegraphics[scale=0.7]{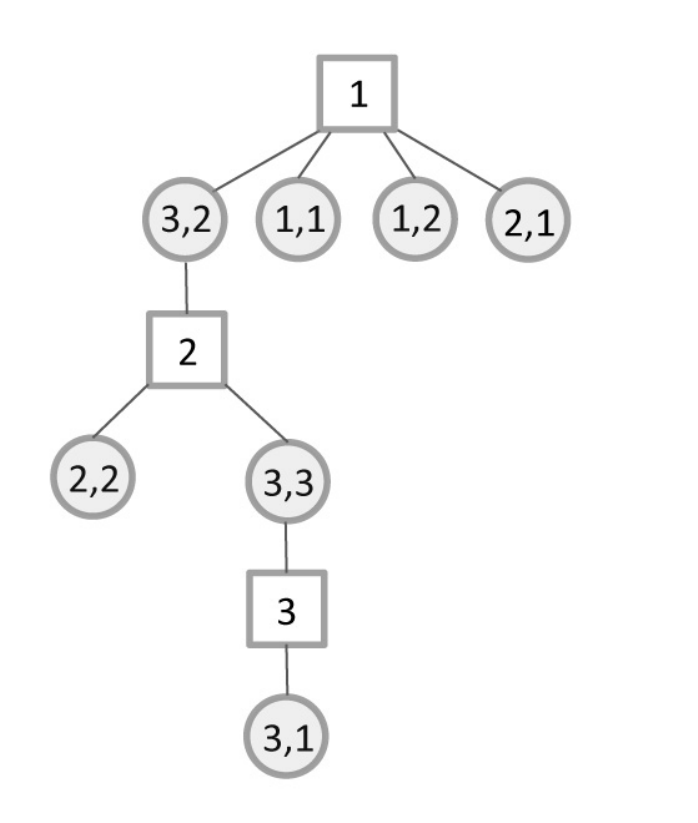}&  
\includegraphics[scale=0.7]{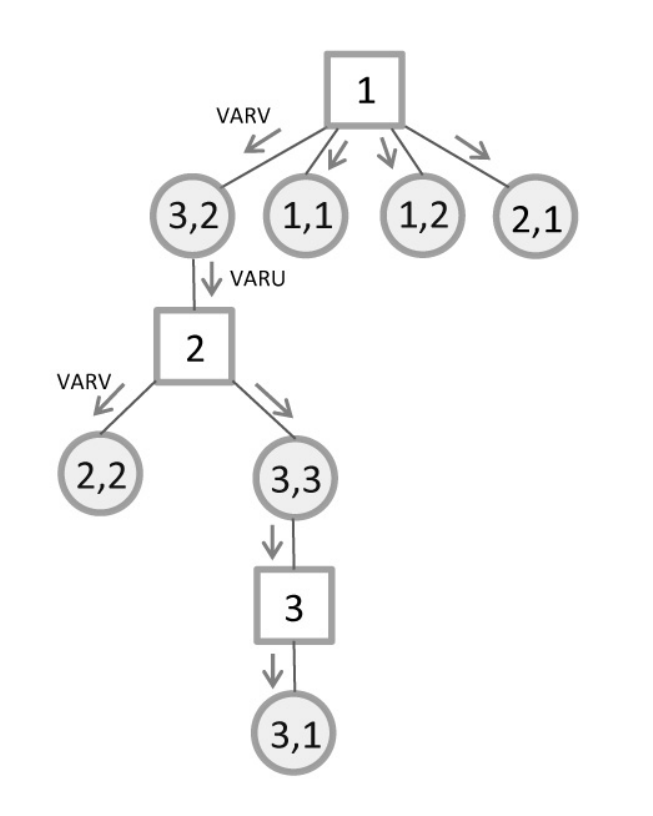} \\
{\bf {\footnotesize(a)}} &
{\bf {\footnotesize(b)}} \\
\end{tabular}

\caption{(a) New tree after a pivot over the shorter cycle; (b) New propagation of dual variables}
\label{fig:5}
\end{figure}

\begin{itemize}

\item Build the initial dual solution tree

In this step, $u_1$ is updated with zero, the corresponding  dual variables $v$ are calculated and sent in {\it Varv} messages. Upon receiving these messages, other processes also  calculate their  dual variables and send these results through {\it Varu} messages, as shown in Figure \ref{fig:2}(a). When all dual variables are calculated, the initial dual solution tree is built, see Figure \ref{fig:2}(b), with the corresponding  dual variables and reduced costs,  as presented in Figures \ref{fig:2}(c) and  \ref{fig:2}(d).

\item Create cycles 

In this step, each server selects its most negative reduced cost edge and sends the message 
{\it Reduced\_cost} to the corresponding server.
This procedure produces candidate cycles, as shown in Figure \ref{fig:3}(a), where three negative costs are selected ($\overline{c}_{3,1}^{1}$,$\overline{c}_{3,2}^{1}$ and $\overline{c}_{3,3}^{1}$). 

\item Select a basic variable to leave the solution

Upon receiving a {\it Reduced\_cost} message, the server sends the {\it Cycle} message along the candidate cycle to select a basic variable to leave the solution, as seen in Figure \ref{fig:3}(b) with reduced cost $\overline{c}_{3,3}^{1}$. The {\it Cycle} message is sent with the value $\theta$ to select the edge flow with the smallest value. Initially, $\theta$ is infinity. The value $-\theta$ or  $+\theta$  is assigned to each edge of the cycle,  alternately. When all edges of the cycle are visited, the smallest flow of the edges is known. Then, an {\it Update} message is sent along the cycle to update the flows of $\theta$ with this smallest value. A basic variable that now has value zero is selected to leave the solution.
It is possible to select until $n$ variables to leave the solution. However, when these candidate cycles have intersections and they are pivoted at the same time, a mutual exclusion problem appears what can produce inconsistent results.

\item  Cancel cycles

 When a conflict is detected, the cycle associated with the
smallest reduced cost is chosen to continue, while the other is canceled.
Suppose that the messages {\it Reduced\_cost} corresponding to the requests 3,1 and  3,3  arrive  in the  server 1 with equal costs,   and then the other {\it Reduced\_cost} message from request 3,2 with a smaller cost reaches it.   In this moment, the server detects a conflict, as shown in Figure \ref{fig:3}(c).
Then, {\it Cancel} messages are sent along the cycle with the highest reduced cost, see Figure \ref{fig:3}(d).
Only after canceling the previous cycle, a new basic variable selection is initiated (see Figure \ref{fig:4}(a)) and the flows are  updated as in Figure \ref{fig:4}(b).

\item Rebuild the dual solution tree

The new tree after the pivot  is rebuilt, as shown in Figure \ref{fig:5}(a), and them the new dual variables are propagated through {\it Varu} and {\it Varv} messages, as presented in Figure \ref{fig:5}(b).

\end{itemize}

All procedures described  before are repeated until it is detected that there is no edge with negative reduced cost. 

\subsubsection{Complexity Analysis}\label{subsubsubsec:complex}

At first we analyse the {\sl DistInit} algorithm that aims to get an initial solution for the TP problem.
Each one of the client servers tries to attend its local requests. In a worst case scenario, 
this procedure can propagate $n$ messages for each local request $m$, resulting in worst case complexities of $O(n.m)$ messages (total number of messages 
sent during the algorithm), $O(n)$ global time (maximum message sequence), and $O(|J_i|)$ local time, where $|J_i|$ is defined
as the number of local requests for each server $i \in I$.

Starting from a solution provided by the {\sl DistInit} method, the {\sl DistTS} initiates the procedure that 
calculates all dual variables by sending $Varv$ and $Varu$ messages, resulting in a complexity of $O(n.m)$ messages. 
Next, each server selects its most negative reduced cost edge and warns the elected corresponding server, which in turn 
broadcasts its election to all other servers. In the worst case, $O(n^2)$ messages are sent to identify candidate cycles to be processed as pivot steps.
The conflicting cycles are canceled with a complexity of $O(n^2)$ messages while each one of the $k$ remaining cycles $(k < n)$ are processed 
and has to re-calculate its dual variables. The new dual variables are propagated through all servers with a total of $O(k.n)$ messages.
The above observations leads to a worst case message complexity for the {\sl DistTS} method of $O(p.n.m)$, where $p$ is the number of rounds 
of the Simplex. 

Considering now the global time of {\sl DistTS}, we note that the maximum message sequence to process the dual variables, identify/cancel cycles 
and broadcast new variables, are all bounded by the height of the dual solution tree, which is $O(n)$. Therefore, we reach a global time complexity of $O(p.n)$.
Moreover, the local time spent by a server is dominated by the process of selecting its most negative reduced cost edge $(O(n))$, 
finding candidate cycles $(O(n))$, or selecting a request to be served  $(O(m))$. Since $O(m) > O(n)$, we achieve a local time complexity of $O(m)$.

\section{Experimental Results}
\label{cap:resultado}

The algorithms described in the previous section were implemented in ANSI C and MPI  (MPICH2) \cite{mpi} for message passing. All experiments were performed (with exclusive access) on a cluster with $42$ nodes, each one with two processors Intel Xeon 
QuadCore $2.66$Hz and $16$Gb of RAM under Linux (Red Hat) $5.3$ operating system. 
The algorithms were tested over the set of instances presented in \cite{tneves} 
for the combined RPP and RRSP problems (both problems are optimized at once).
These instances were adapted for the RRSP by fixing the locations of the contents 
on the servers. 
These instances are divided into three classes: easy, medium and hard. The instances from the easy class
are not considered in this work.
There are twenty instances for each classes, five for each value of $n \in \{10,20,30,50\}$. The
average number of requests per server is 70. The entire benchmark is available for download in \cite{labic}.

\subsection{Comparison of DistTS and the Sequential Version}

The first experiment, reported in Table \ref{tab:result}, is a comparison of the new distributed heuristic
{\sl DistInit} with the sequential Minimum Cost Method (MCM), in terms of solution quality. The instances
are identified by the label $n\_y$, instances with $y$ ranging from 6 to 10 belong to class medium, those
where $y$ is between 11 and 15 are from class hard. The second and third columns present the number of requests and the value of an optimal solution, respectively. The next two columns are the value of the solution found by
MCM and the corresponding percentual difference to the optimal. The next columns are the average
value of the solutions found by 10 executions of {\sl DistInit} (since this algorithm is not deterministic), the corresponding average difference to the optimal and the standard deviation. {\sl DistInit} performed a little better than MCM. 
Some very large solutions values indicate the use of some artificial variables, so those
solutions are not really feasible. MCM could not find a feasible solution on 17 instances, this happened 14 times
with {\sl DistInit}. Considering only the 23 instances where both algorithms could find a feasible solution, the average
difference to the optimal is 2.4\% for MCM and 2.2\% for {\sl DistInit}. The small values in the last colummn indicate that 
{\sl DistInit} found very similar solutions on its different executions of the same instance.
However, it should be stressed that the tests were performed in a ``well-behaved'' environment. 
Processor loads and message delays would be much more unpredictable in a real
CDN environment.
  
\begin{table}[htp]
\scriptsize
\caption{\label{tab:result}Heuristic solution: {\sl DistInit} x MCM}
\begin{center}	
\begin{tabular}{r|r|r|rr|rrr}
\hline

\multirow{2}{*}{Instance}   & 	\multirow{2}{*}{Requests} &\multirow{2}{*}{Opt}	& \multicolumn{2}{c|}{MCM}	& \multicolumn{3}{c}{DistInit}  \\
\cline{4-8}
	                        &		                        &                   &	Value	&  \%Opt &  Value 	& \%Opt	 & SD	\\
	\hline
10\_6	&	627	&	471,763	&	\textbf{471,763}	&	0.0	&	\textbf{471,763}	&	0.0	&	0.0	\\	
10\_7	&	675	&	306,227	&	12,055,281	&	3,836.7	&	\textbf{306,227}	&	0.0	&	0.0	\\	
10\_8	&	641	&	378,231	&	380,743	&	0.7	&	\textbf{379,416}	&	0.3	&	0.0	\\	
10\_9	&	659	&	461,161	&	360,134,315	&	77,992.9	&	\textbf{66,494,292}	&	14,318.9	&	57.5	\\	
10\_10	&	649	&	501,082	&	\textbf{516,506}	&	3.1	&	516,538	&	3.1	&	1.1	\\	
10\_11	&	627	&	471,763	&	\textbf{471,763}	&	0.0	&	\textbf{471763}	&	0.0	&	0.0	\\	
10\_12	&	675	&	316,065	&	104,204,534	&	32,869.3	&	\textbf{93,844,995}	&	29,591.7	&	7.6	\\	
10\_13	&	641	&	378,231	&	380,743	&	0.7	&	\textbf{379,416}	&	0.3	&	0.0	\\	
10\_14	&	659	&	461,161	&	360,134,315	&	77,992.9	&	\textbf{104,635,498}	&	22,589.6	&	54.3	\\	
10\_15	&	649	&	501,082	&	\textbf{516,506}	&	3.1	&	519,473	&	3.7	&	0.8	\\	
\hline																
20\_6	&	1,289	&	1,434,570	&	298,622,394	&	20,716.2	&	\textbf{69,422,807}	&	4,739.3	&	106.5	\\	
20\_7	&	1,356	&	916,306	&	974,232	&	6.3	&	\textbf{950,868}	&	3.8	&	0.6	\\	
20\_8	&	1,314	&	1,158,373	&	252,787,670	&	21,722.6	&	\textbf{186,742,182}	&	16,021.1	&	34.6	\\	
20\_9	&	1,352	&	1,160,991	&	814,687,624	&	70,071.7	&	\textbf{542,353,658}	&	46,614.7	&	18.8	\\	
20\_10	&	1,367	&	853,256	&	864,902	&	1.4	&	\textbf{864,245}	&	1.3	&	0.1	\\	
20\_11	&	1,289	&	1,434,570	&	298,622,394	&	20,716.2	&	\textbf{90,560,889}	&	6,212.8	&	80.4	\\	
20\_12	&	1,356	&	916,306	&	974,232	&	6.3	&	\textbf{951,267}	&	3.8	&	0.7	\\	
20\_13	&	1,314	&	1,158,373	&	252,787,670	&	21,722.6	&	\textbf{181,488,041}	&	15,567.5	&	19.7	\\	
20\_14	&	1,352	&	1,160,991	&	814,687,624	&	70,071.7	&	\textbf{481,505,793}	&	41,373.7	&	22.6	\\	
20\_15	&	1,367	&	839,076	&	845,561	&	0.8	&	\textbf{845,503}	&	0.8	&	0.2	\\	
\hline																
30\_6	&	2,007	&	1,443,887	&	\textbf{1,452,805}	&	0.6	&	1,471,337	&	1.9	&	0.3	\\	
30\_7	&	1,963	&	1,280,967	&	\textbf{1,299,447}	&	1.4	&	1,311,361	&	2.4	&	0.2	\\	
30\_8	&	2,021	&	1,132,309	&	1,187,076	&	4.8	&	\textbf{1,168,246}	&	3.2	&	1.6	\\	
30\_9	&	1,991	&	1,169,035	&	\textbf{1,219,235}	&	4.3	&	1,226,495	&	4.9	&	0.4	\\	
30\_10	&	1,998	&	1,150,971	&	111,066,244	&	9,549.8	&	\textbf{87,874,822}	&	7,534.8	&	50.7	\\	
30\_11	&	2,007	&	1,458,771	&	\textbf{1,467,689}	&	0.6	&	1,487,350	&	2.0	&	0.2	\\	
30\_12	&	1,963	&	1,280,967	&	\textbf{1,299,447}	&	1.4	&	1,308,424	&	2.1	&	0.4	\\	
30\_13	&	2,021	&	1,132,309	&	1,187,076	&	4.8	&	\textbf{1,168,002}	&	3.2	&	1.5	\\	
30\_14	&	1,991	&	1,182,625	&	\textbf{1,232,757}	&	4.2	&	1,241,409	&	5.0	&	0.3	\\	
30\_15	&	1,998	&	1,150,971	&	111,066,244	&	9,549.8	&	\textbf{68,044,774}	&	5,811.9	&	113.0	\\	
\hline																
50\_6	&	3,391	&	2,223,512	&	232,991,437	&	10,378.5	&	\textbf{2,326,645}	&	4.6	&	0.9	\\	
50\_7	&	3,329	&	1,655,212	&	1,674,253	&	1.2	&	\textbf{1,670,467}	&	0.9	&	0.2	\\	
50\_8	&	3,214	&	3,065,126	&	\textbf{81,444,220}	&	2,557.1	&	100,397,235	&	3,175.5	&	68.0	\\	
50\_9	&	3,303	&	3,029,901	&	\textbf{3,111,283}	&	2.7	&	3,117,665	&	2.9	&	0.4	\\	
50\_10	&	3,295	&	2,196,471	&	2,250,427	&	2.5	&	\textbf{2,227,284}	&	1.4	&	0.5	\\	
50\_11	&	3,391	&	2,223,512	&	232,991,437	&	10,378.5	&	\textbf{2,333,775}	&	5.0	&	0.7	\\	
50\_12	&	3,329	&	1,655,212	&	1,674,253	&	1.2	&	\textbf{1,668,489}	&	0.8	&	0.2	\\	
50\_13	&	3,314	&	3,065,126	&	\textbf{81,444,220}	&	2,557.1	&	162,122,968	&	5,189.3	&	35.3	\\	
50\_14	&	3,303	&	3,011,459	&	163,208,074	&	5,319.6	&	\textbf{14,219,738}	&	372.2	&	126.4	\\	
50\_15	&	3,295	&	2,196,403	&	2,223,243	&	1.2	&	\textbf{2,217,503}	&	1.0	&	0.2	\\	
\hline



\end{tabular}
\end{center}
\end{table} 

The second experiment, reported in Table \ref{tab:result1}, compare the sequential Transportation Simplex method, initialized with
the MCM, with the distributed Transportation Simplex method ({\sl DistTS}), initialized with {\sl DistInit}. First, there is a comparison of their performance on finding the first truly feasible solution, without using artificial variables. In some practical contexts, one just needs a feasible solution, so the methods can be stopped at that point. The table shows, for each method, the value of the first solution and the number of pivot steps necessary to find it. If the initial heuristic already produces a feasible solution, the number of pivots is zero. The number of steps necessary to find the optimal solution is also given. 
It can be seen that the performance of both methods is similar in terms of pivots required to find an optimal solution. This means that the fact that some pivots are being made in parallel is not affecting the algorithm. However, the level
of parallelism observed in this experiment was small, it is rare that more than 3 pivots can be performed in parallel. 
In those instances, the candidate cycles are composed of too many servers and requests, which results in lots of intersections and, consequently,  poor parallelism. Its due to the characteristics of the instances, which were randomly generated in such a way that the requests for a given content are evenly distributed among the servers. Therefore, that are no natural clusters of requests/contents that can be optimized in parallel.

\noindent
\begin{table}[htp]
\scriptsize
\setlength{\tabcolsep}{1.5mm}
\caption{\label{tab:result1}Transportation Simplex: Sequential x Distributed}
\begin{center}	
\begin{tabular}{r|rr|r|rr|r}
\hline
\multirow{4}{*}{Instance}&  \multicolumn{3}{c|}{Sequential} &\multicolumn{3}{c}{Distributed}  \\
\cline{2-7}
& \multicolumn{2}{c|}{Feasible} & \multicolumn{1}{c|}{Final} & \multicolumn{2}{c|}{Feasible} & \multicolumn{1}{c}{Final} \\
& \multicolumn{2}{c|}{Solution} & \multicolumn{1}{c|}{Solution} & \multicolumn{2}{c|}{Solution} & \multicolumn{1}{c}{Solution}\\
\cline{2-7}
	&\multirow{1}{*}{Solution} 	&	Pivots	&	Pivots	&	\multirow{1}{*}{Solution}	&  Pivots 	&		Pivots  \\
	\hline
10\_6	&	471,763	&	0	&	0	&	471,763	&	0	&	0	\\		
10\_7	&	306,303	&	1	&	2	&	306,227	&	0	&	0	\\		
10\_8	&	380,743	&	0	&	4	&	380,365	&	0	&	1	\\		
10\_9	&	469,533	&	18	&	23	&	491,423	&	2	&	20	\\		
10\_10	&	516,506	&	0	&	10	&	526,855	&	0	&	17	\\		
10\_11	&	471,763	&	0	&	0	&	471,763	&	0	&	0	\\		
10\_12	&	316,443	&	5	&	6	&	316,874	&	2	&	3	\\		
10\_13	&	380,743	&	0	&	4	&	380,615	&	0	&	1	\\		
10\_14	&	469,533	&	18	&	23	&	491,653	&	4	&	22	\\		
10\_15	&	516,506	&	0	&	10	&	527,942	&	0	&	21	\\		
\hline															
20\_6	&	1,465,556	&	17	&	52	&	1,555,328	&	5	&	46	\\		
20\_7	&	974,232	&	0	&	86	&	938,719	&	0	&	77	\\		
20\_8	&	1,165,685	&	12	&	26	&	1,385,258	&	12	&	32	\\		
20\_9	&	1,162,230	&	57	&	74	&	1,170,782	&	43	&	86	\\		
20\_10	&	864,902	&	0	&	17	&	866,121	&	0	&	21	\\		
20\_11	&	1,465,556	&	17	&	52	&	1,469,270	&	6	&	45	\\		
20\_12	&	974,232	&	0	&	86	&	937,778	&	0	&	79	\\		
20\_13	&	1,165,685	&	12	&	26	&	1,225,276	&	11	&	35	\\		
20\_14	&	1,162,230	&	57	&	74	&	1,169,135	&	43	&	81	\\		
20\_15	&	845,561	&	0	&	12	&	848,275	&	0	&	12	\\		
\hline															
30\_6	&	1,452,805	&	0	&	33	&	1,471,605	&	0	&	42	\\		
30\_7	&	1,299,447	&	0	&	36	&	1,299,497	&	0	&	49	\\		
30\_8	&	1,187,076	&	0	&	63	&	1,166,714	&	0	&	60	\\		
30\_9	&	1,219,235	&	0	&	78	&	1,222,032	&	0	&	94	\\		
30\_10	&	1,178,768	&	9	&	46	&	1,175,020	&	6	&	53	\\		
30\_11	&	1,467,689	&	0	&	33	&	1,486,846	&	0	&	42	\\		
30\_12	&	1,299,447	&	0	&	36	&	1,299,569	&	0	&	47	\\		
30\_13	&	1,187,076	&	0	&	63	&	1,169,751	&	0	&	60	\\		
30\_14	&	1,232,757	&	0	&	76	&	1,235,904	&	0	&	96	\\		
30\_15	&	1,178,768	&	9	&	46	&	1,171,555	&	8	&	49	\\		
\hline															
50\_6	&	2,351,415	&	25	&	170	&	2,322,403	&	0	&	149	\\		
50\_7	&	1,674,253	&	0	&	32	&	1,666,277	&	0	&	29	\\		
50\_8	&	3,149,847	&	7	&	121	&	3,144,390	&	12	&	143	\\		
50\_9	&	3,111,283	&	0	&	108	&	3,123,216	&	0	&	103	\\		
50\_10	&	2,250,427	&	0	&	60	&	2,236,384	&	0	&	46	\\		
50\_11	&	2,351,415	&	25	&	170	&	2,308,857	&	0	&	149	\\		
50\_12	&	1,674,253	&	0	&	32	&	1,664,698	&	0	&	27	\\		
50\_13	&	3,149,847	&	7	&	121	&	3,197,215	&	13	&	154	\\		
50\_14	&	3,078,389	&	11	&	124	&	3,201,781	&	1	&	109	\\		
50\_15	&	2,223,243	&	0	&	38	&	2,222,120	&	0	&	40	\\		
\hline															
Average	&	1,331,579	&	8	&	52	&	1,342,931	&	4	&	54	\\		
\hline

\end{tabular}
\end{center}
\end{table} 

\subsection{Comparison of DistTS and AuctionTP}

The third experiment, reported in Table \ref{tab:result2}, compares the {\sl DistTS} and {\sl AuctionTP} methods. 
The first and second columns present the name and the optimal solution of the instances. The next four columns contain the first feasible solution, the time in seconds to find it, the total time and the number of exchanged messages by the {\sl DistTS}. The last four columns show the same information for the {\sl AuctionTP}.

It is possible to observe that for instances with 10 servers both algorithms present similar execution time. As the number of servers increases, the {\sl AuctionTP} execution time also grows.
It occurs because the bidding and assignment steps, executed several times until the optimal solution is found,  are separated logically by synchronization points. In the distributed algorithm it means that a server can not progress  to the next bidding step without receiving an acknowledge message from each other destination  to which it sent a bidding message. On its turns, without receiving all bidding messages,  destination processes do not update their prices and send the corresponding  acknowledge messages.
Clearly, the time spent with synchronization increases with the number of servers.


\noindent
\begin{table}[htp]
\scriptsize
\setlength{\tabcolsep}{1.5mm}
\caption{\label{tab:result2}DistTS x AuctionTP}
\begin{center}	
\begin{tabular}{r|r|rrrr|rrrr}
\hline																			
\multirow{3}{*}{Instance}	&	\multirow{3}{*}{Opt}	&	\multicolumn{4}{c|}{DistTS}	&							\multicolumn{4}{c}{AuctionTP}							\\
\cline{3-10}																			
	&		&	Feasible 	&	Feasible 	&	Total 	&	Total	&	Feasible 	&	Feasible 	&	Total 	&	Total 	\\
	&		&	Solution	&	Time(s)	&	Time(s)	&	Messages	&	Solution	&	Time(s)	&	Time(s)	&	Messages	\\
\hline																			
10\_6	&	471,763	&	471,763	&	0.00	&	0.00	&	953	&	471,763	&	0.02	&	1.09	&	366,600	\\
10\_7	&	306,227	&	306,227	&	0.00	&	0.01	&	1,424	&	306,227	&	0.04	&	1.07	&	357,500	\\
10\_8	&	378,231	&	380,365	&	0.00	&	0.01	&	2,298	&	378,231	&	0.04	&	1.17	&	370,500	\\
10\_9	&	461,161	&	491,423	&	0.00	&	0.08	&	10,215	&	461,161	&	0.04	&	0.88	&	289,900	\\
10\_10	&	501,082	&	526,855	&	0.00	&	0.09	&	11,762	&	501,082	&	0.03	&	1.12	&	357,500	\\
10\_11	&	471,763	&	471,763	&	0.00	&	0.00	&	953	&	471,763	&	0.02	&	1.10	&	366,600	\\
10\_12	&	316,065	&	316,874	&	0.01	&	0.02	&	3,069	&	316,065	&	0.04	&	1.07	&	357,500	\\
10\_13	&	378,231	&	380,615	&	0.00	&	0.01	&	2,558	&	378,231	&	0.04	&	1.14	&	370,500	\\
10\_14	&	461,161	&	491,653	&	0.00	&	0.08	&	10,210	&	461,161	&	0.04	&	0.90	&	289,900	\\
10\_15	&	501,082	&	527,942	&	0.00	&	0.10	&	12,548	&	501,082	&	0.03	&	1.11	&	357,500	\\
\hline																			
20\_6	&	1,434,570	&	1,555,328	&	0.13	&	0.69	&	93,428	&	1,434,570	&	16.20	&	170.86	&	5,617,416	\\
20\_7	&	916,306	&	938,719	&	0.02	&	0.79	&	100,124	&	916,306	&	85.75	&	167.44	&	5,501,080	\\
20\_8	&	1,158,373	&	1,385,258	&	0.23	&	0.44	&	58,628	&	1,158,373	&	15.71	&	233.78	&	6,177,400	\\
20\_9	&	1,160,991	&	1,170,782	&	0.68	&	1.09	&	141,779	&	1,160,991	&	10.82	&	195.10	&	6,275,040	\\
20\_10	&	853,256	&	866,121	&	0.01	&	0.32	&	41,636	&	853,256	&	2.83	&	164.35	&	5,299,200	\\
20\_11	&	1,434,570	&	1,469,270	&	0.13	&	0.71	&	94,914	&	1,434,570	&	16.81	&	171.44	&	5,617,416	\\
20\_12	&	916,306	&	937,778	&	0.01	&	0.65	&	86,582	&	916,306	&	86.18	&	168.40	&	5,497,504	\\
20\_13	&	1,158,373	&	1,225,276	&	0.20	&	0.41	&	55,179	&	1,158,373	&	15.64	&	235.57	&	6,177,400	\\
20\_14	&	1,160,991	&	1,169,135	&	0.82	&	1.23	&	148,894	&	1,160,991	&	10.93	&	195.71	&	6,280,480	\\
20\_15	&	839,076	&	848,275	&	0.01	&	0.19	&	25,287	&	839,076	&	2.90	&	165.70	&	5,301,500	\\
\hline																			
30\_6	&	1,443,887	&	1,471,605	&	0.08	&	12.71	&	185,099	&	1,443,887	&	26.44	&	535.24	&	31,672,992	\\
30\_7	&	1,280,967	&	1,299,497	&	0.27	&	12.65	&	152,457	&	1,280,967	&	13.35	&	468.23	&	27,845,568	\\
30\_8	&	1,132,309	&	1,166,714	&	0.13	&	19.90	&	206,759	&	1,132,309	&	8.00	&	484.28	&	28,915,614	\\
30\_9	&	1,169,035	&	1,222,032	&	0.05	&	24.74	&	272,909	&	1,169,035	&	7.73	&	502.68	&	29,040,348	\\
30\_10	&	1,150,971	&	1,175,020	&	2.72	&	13.44	&	181,513	&	1,150,971	&	47.75	&	498.32	&	29,718,000	\\
30\_11	&	1,458,771	&	1,486,846	&	0.07	&	11.85	&	178,703	&	1,458,771	&	25.81	&	535.75	&	31,692,180	\\
30\_12	&	1,280,967	&	1,299,569	&	0.34	&	12.32	&	148,595	&	1,280,967	&	13.41	&	479.81	&	28,547,466	\\
30\_13	&	1,132,309	&	1,169,751	&	0.23	&	20.31	&	214,672	&	1,132,309	&	7.95	&	485.60	&	28,909,776	\\
30\_14	&	1,182,625	&	1,235,904	&	0.07	&	26.18	&	288,581	&	1,182,625	&	7.62	&	500.15	&	29,034,486	\\
30\_15	&	1,150,971	&	1,171,555	&	2.52	&	12.48	&	164,366	&	1,150,971	&	47.95	&	496.07	&	29,730,000	\\
\hline																			
50\_6	&	2,223,512	&	2,322,403	&	1.13	&	165.27	&	1,318,262	&	2,223,512	&	157.50	&	1,712.05	&	251,921,000	\\
50\_7	&	1,655,212	&	1,666,277	&	1.06	&	34.85	&	245,676	&	1,655,212	&	84.35	&	1,520.94	&	253,663,800	\\
50\_8	&	3,065,126	&	3,144,390	&	6.90	&	138.37	&	1,279,083	&	3,065,126	&	90.36	&	1,580.04	&	245,158,910	\\
50\_9	&	3,029,901	&	3,123,216	&	1.05	&	125.19	&	949,613	&	3,029,901	&	135.05	&	1,673.37	&	257,829,200	\\
50\_10	&	2,196,471	&	2,236,384	&	0.90	&	57.01	&	455,113	&	2,196,471	&	164.02	&	1,676.71	&	245,177,400	\\
50\_11	&	2,223,512	&	2,308,857	&	1.08	&	160.70	&	1,298,531	&	2,223,512	&	157.29	&	1,693.02	&	251,864,000	\\
50\_12	&	1,655,212	&	1,664,698	&	1.07	&	22.85	&	165,867	&	1,655,212	&	84.63	&	1,522.36	&	253,663,800	\\
50\_13	&	3,065,126	&	3,197,215	&	21.02	&	152.57	&	1,427,140	&	3,065,126	&	90.56	&	1,549.89	&	240,678,250	\\
50\_14	&	3,011,459	&	3,201,781	&	1.99	&	135.31	&	1,001,733	&	3,011,459	&	26.13	&	1,517.02	&	241,516,500	\\
50\_15	&	2,196,403	&	2,222,120	&	1.04	&	49.35	&	398,012	&	2,196,403	&	162.15	&	1,626.70	&	251,892,500	\\
\hline																			
Average	&	1,299,608	&	1,342,931	&	1.15	&	30.37	&	285,878	&	1,299,608	&	40.55	&	573.43	&	71,242,506	\\
\hline

\end{tabular}
\end{center}
\end{table}

\section{Final Remarks}
\label{conclusao}

This paper tackled a known problem of CDNs, the Request Routing System
Problem, as a Transportation Problem. In the related literature, simplex
and auction based approaches are usually presented to solve the TP in
shared memory parallel machines. However, as CDN information is spread
out on servers that communicate only  by message passing,  new parallel
algorithms adjusted to this distributed scenario have been  proposed here.

Our main contribution  was the design and implementation of a distributed
transportation Simplex algorithm adapted to solve the RRSP. It includes a
distributed heuristic for finding an initial solution that it is
interesting on its own.
Furthermore, for comparison purposes, we also developed a distributed
Auction algorithm based on the shared memory parallel version of
\cite{bertsekas}.

Experiments with instances adapted from the literature were performed. The
results pointed that the $DistInit$ and $DistTS$ algorithms have a
performance similar to their sequential counterparts, in spite of not
requiring global information about the content requests. However, only a
limited amount of parallelism was obtained in the tested instances.
This was majorly due to the fact that in tested instances there were few
natural clusters of requests/contents that could be processed in parallel.
Moreover, we also verified that $DistTS$ outperformed the {\sl AuctionTP}
in the largest instances, showing that the synchronization points of {\sl
AuctionTP} introduced a high overhead in a distributed environment. Future work will concentrate on the investigation of the parameter $\varepsilon$ in order to improve the convergence time of {\sl AuctionTP}.

\bibliographystyle{plain}
\bibliography{ref}\label{sec:refs}

\end{document}